\documentclass[10pt,letterpaper]{article}

\usepackage{cogsci}
\cogscifinalcopy 

\usepackage{graphicx}
\usepackage{amsmath,amssymb,amsthm,amsfonts}
\usepackage{acronym}
\usepackage{enumitem}
\usepackage{xspace}
\usepackage{cancel}
\usepackage[normalem]{ulem}

\usepackage[switch]{lineno}
\usepackage{url}
\usepackage{multirow}
\usepackage{booktabs}
\usepackage{mathrsfs}
\usepackage{arydshln}

\usepackage{array}
\usepackage{colortbl}
\usepackage{pifont}

\makeatletter
\def\@makefntext#1{%
  \hb@xt@ 1.8em{\hss}#1
}

\DeclareRobustCommand\onedot{\futurelet\@let@token\@onedot}
\def\@onedot{\ifx\@let@token.\else.\null\fi\xspace}
\def\eg{\emph{e.g}\onedot}

\def\ie{\emph{i.e}\onedot}

\def\etc{\emph{etc}\onedot}

\makeatother

\acrodef{nlp}[NLP]{natural language processing}
\acrodef{plm}[PLM]{pretrained language model}
\acrodef{sota}[SOTA]{state-of-the-art}
\acrodef{bs}[BS]{Beam Search}
\acrodef{mhs}[MHS]{Metropolis-Hastings Sampling}
\acrodef{hs}[HS]{Hybrid Search}
\acrodef{uas}[UAS]{unlabeled attachment score}
\acrodef{dda}[DDA]{Directed Dependency Accuracy}
\acrodef{sota}[SOTA]{state-of-the-art}
\acrodef{pos}[POS]{part-of-speech}

\usepackage{pifont}
\usepackage{hyperref}
\usepackage{pslatex}
\usepackage{apacite}

\usepackage{makecell} 
\usepackage{booktabs} 

\usepackage{float} 
\usepackage{caption} 
\usepackage[misc]{ifsym}

\newcommand{\simulator}{{\fontfamily{lmtt}\selectfont PersonalityScanner}\xspace}

\title{PersonalityScanner: Exploring the Validity of Personality Assessment Based on Multimodal Signals in Virtual Reality}

\author{
\large \bf Xintong Zhang \textsuperscript{\textrm1,2},
Di Lu \textsuperscript{\textrm3},
Huiqi Hu \textsuperscript{\textrm1},
Nan Jiang \textsuperscript{\textrm2,4},
Xianhao Yu \textsuperscript{\textrm1}, \\
\large \bf Jinan Xu \textsuperscript{\textrm1}, 
Yujia Peng \textsuperscript{\textrm2,3,4},
Qing Li\textsuperscript{\textrm2}, 
Wenjuan Han$^{\textrm{\Letter}}$\thanks{\Letter \, Corresponding author.} \\
 \textsuperscript{\textrm 1}Beijing Key Lab of Traffic Data Analysis and Mining, Beijing Jiaotong University, Beijing, China \\
  \textsuperscript{\textrm 2}State Key Laboratory of General Artificial Intelligence, Beijing Institute for General Artificial Intelligence (BIGAI) \\
 \textsuperscript{\textrm 3}School of Psychological and Cognitive Sciences and Beijing Key Laboratory of Behavior and Mental Health, Peking University\\
  \textsuperscript{\textrm 4}Institute for Artificial Intelligence, Peking University\\
}

\begin{document}

\maketitle

\begin{abstract}
Human cognition significantly influences expressed behavior and is intrinsically tied to authentic personality traits. Personality assessment plays a pivotal role in various fields, including psychology, education, social media, \etc. However, traditional self-report questionnaires can only provide data based on what individuals are willing and able to disclose, thereby lacking objective. Moreover, automated measurements and peer assessments demand significant human effort and resources. In this paper, given the advantages of the Virtual Reality (VR) technique, we develop a VR simulator --- \simulator, to stimulate cognitive processes and simulate daily behaviors based on an immersive and interactive simulation environment, in which participants carry out a battery of engaging tasks that formulate a natural story of first-day at work. Through this simulator, we collect a synchronous multi-modal dataset with ten modalities, including first/third-person video, audio, text, eye tracking, facial microexpression, pose, depth data, log, and inertial measurement unit. By systematically examining the contributions of different modalities on revealing personality, we demonstrate the superior performance and effectiveness of \simulator.


\end{abstract}


The cognitive process encompasses perception, memory, thinking, judgment, and problem-solving. These psychological processes influence how individuals interpret and understand the world around them, thereby affecting their responses and behaviors~\cite{kogan1964risk, schaie2004seattle}, including body posture, facial expressions, and voice. For instance, an optimistic cognitive orientation may lead a person to take proactive actions when facing challenges. The mode of cognition affects the choice of behavior, and these behavioral patterns accumulate over time, forming the unique personality traits of an individual. Personality refers to the enduring patterns of behavior, emotion, and thought processes of an individual
 and has been studied extensively in fields of psychology, education, and social media, \etc \cite{costa2008revised,kalat2016introduction}.
 Personality assessments help individuals better understand their own strengths, weaknesses, motivations, and preferences. This self-awareness can be beneficial in a variety of areas, such as making better decisions, career planning, relationship building, and personal development~\cite{caspi2005personality,weiner2017handbook}.

\begin{figure*}[!t]
    \centering
    \includegraphics[width=0.95\linewidth]{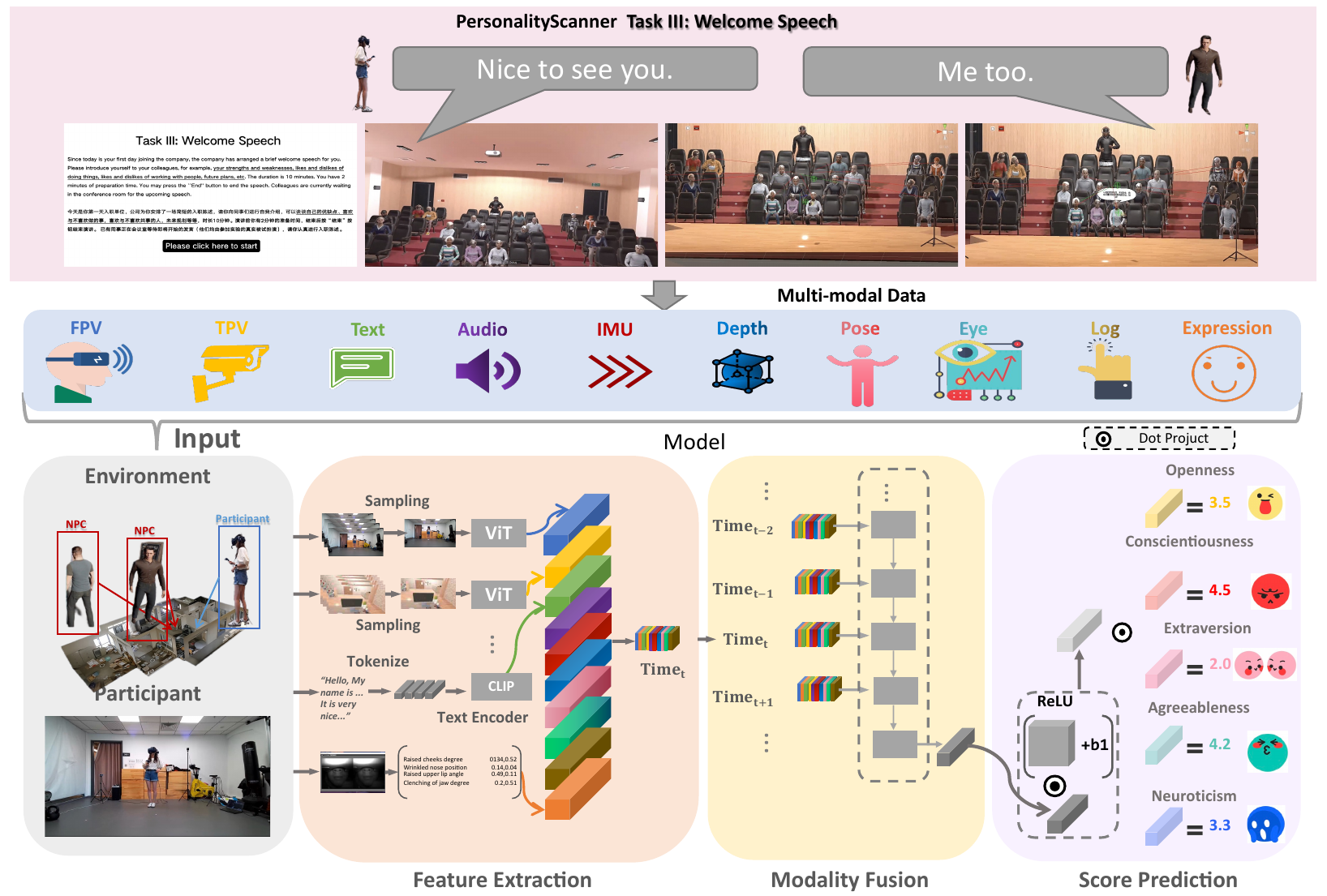}
    \caption{A overview of \simulator simulator and our model to estimate personality.}
    \label{fig:motivation}
\end{figure*}

Traditional personality assessments primarily rely on self-reporting of personality traits~\cite{weiner2017handbook} and peer evaluations~\cite{mcneill2000face,oosterhof2008functional}.
However, the traditional approaches can only yield information that people are able and willing to report rather than real responses controlled by their cognition and suffer from a lack of objective, multi-dimensional, and automatic measurements. To overcome these limitations, the field of computational personality assessment has emerged~\cite{ilmini2017computational, palmero2021context}
leveraging computational techniques to predict personality traits more comprehensively and objectively. 
Among computational approaches, virtual reality (VR) based approaches recently attracted widespread interest~\cite{majumder2017deep,iacobelli2011large}, due to potential advantages of VR (\eg, experimental design' ease of creation, control, and standardization~\cite{maharani2023relationship}, safe environment with minimal exposure to hazardous situations~\cite{maharani2023relationship}, effectively recording multi-modal sensor information~\cite{xie2021review}). 
For instance, \cite{katifori2022exploring} explored how personality traits influence participants' performance of object manipulation tasks in VR, specifically the task of selecting and moving a cube inside a hole. 
They proposed several related hypotheses for verification, such as participants with higher levels of openness are more likely to exhibit increased hand movements.
Similarly, \cite{maharani2023relationship} asked participants to perform asphalt compaction tasks by manually steering the wheel and joystick to reveal the relationship between personality and specific task performance. However, both studies focus solely on basic object manipulation tasks and only consider hand movements, neglecting other body movements and interactions with others, which can provide valuable insights~\cite{sackett1998ability1}.

In this work, we aim to utilize VR technology to identify human cognitive-controlled behaviors during the execution of complex daily tasks to reveal personality traits.
To the best of our knowledge, no prior research has explored the connection between personality traits and complex daily-life activities within an interactive VR simulator that allows participants to freely move their whole bodies, engage in open-ended conversations with non-playable characters (NPCs), and interact with the environment.
To address this gap, we develop \simulator, an interactive VR simulator with advanced NPCs, personality-sensitive tasks, and versatile technological features including real-time data processing, high frame rates, modular VR control, and user-friendly 3D map editing. In \simulator, participants experience their first day at a company, interact with NPCs, and complete tasks (as seen in Figure~\ref{fig:motivation}). We collect multi-modal data during the experiments, which we later use to train a model that maps behavioral performance to personality traits. This model enables us to predict participants' personalities when enough data is available.

In summary, our study makes three significant contributions: (1) We introduce \simulator, an interactive and immersive VR simulator, capable of simulating everyday cognitive environments and social interactions. (2) We collect a synchronous multimodal dataset using the VR simulator, encompassing ten modalities. 
This dataset serves as a new benchmark for evaluating personality assessment performance. (3) Our comprehensive study of different modality types proves the superiority of \simulator and our collected dataset in estimating personality traits.

\begin{figure}[!t]
    \flushleft
        \centering
    \includegraphics[width=1\linewidth]{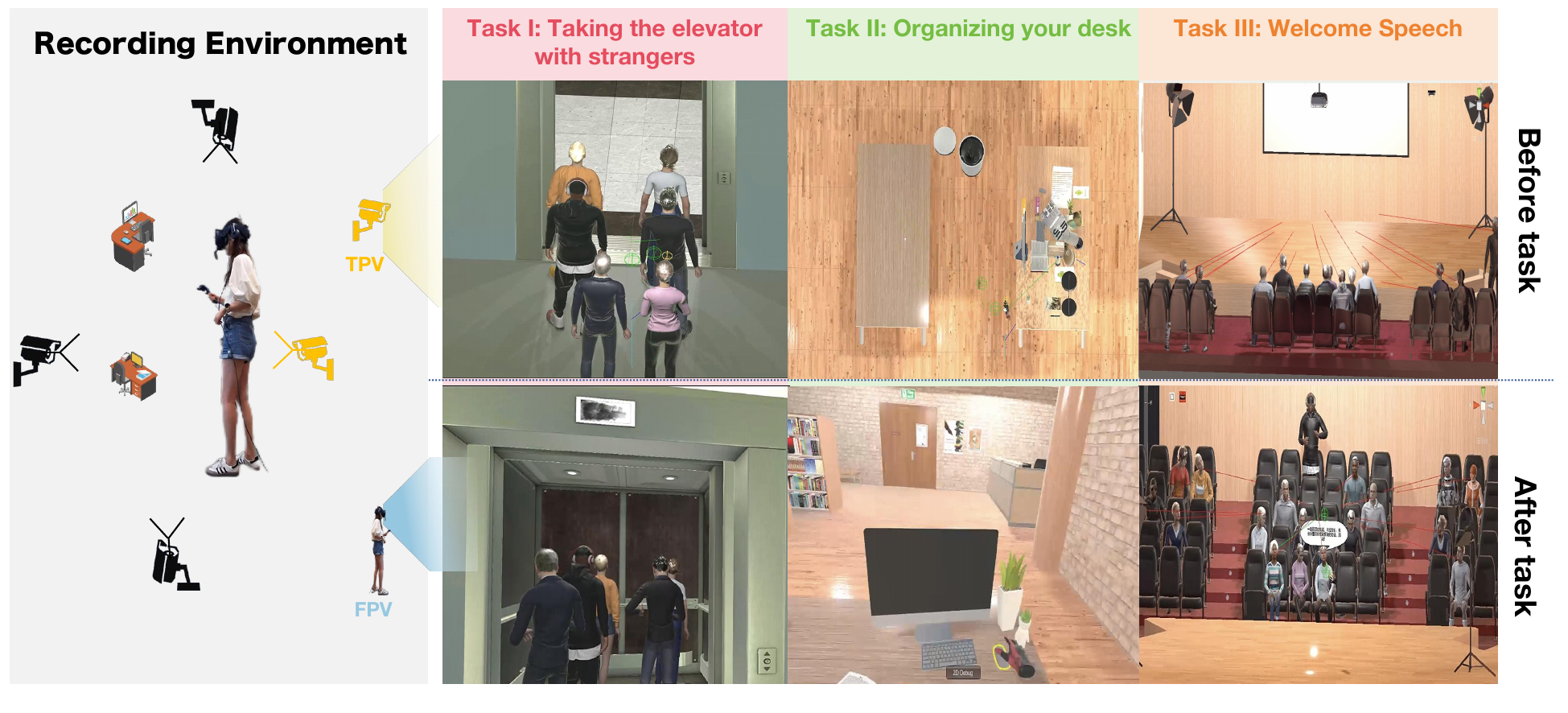}
   \caption{Three tasks of \simulator simulator.}
    \label{fig:simulator}
\end{figure}

\section{Related Work}\label{sec:related_work}
\subsection{Human Cognition and Personality Traits}
According to research findings~\cite{kumari2004personality}, personality traits such as extraversion, conscientiousness, and neuroticism are closely associated with human brain structure and gray matter activity, which serve as critical determinants and controllers of human cognitive processes. 
Previous studies in psychology have also frequently asserted that individuals' personalities are intimately connected with their cognitive processes in various everyday activities, such as adventurous behaviors~\cite{kogan1964risk} and creativity~\cite{mccrae1993openness}.
Importantly, ~\cite{schaie2004seattle} shows that the link between human cognition and personality is stable, sometimes lasting over 35 years. This suggests that cognition is a reliable indicator of personality.

The widely-used Big Five Personality Factors (Big Five)~\cite{costa1999five,mccrae1997personality} 
is used as our theoretical foundation for assessing personality. It categorizes human personality using five key traits: 
\begin{itemize}[leftmargin=*,noitemsep,nolistsep]
    \item \textbf{\underline{O}penness}: artistic, curious, imaginative, insightful, and original with wide interests.
    \item \textbf{\underline{C}onscientiousness}: efficient, organized, planful, reliable, responsible, and thorough.
    \item \textbf{\underline{E}xtraversion}: active, assertive, energetic, enthusiastic, outgoing, and talkative.
    \item \textbf{\underline{A}greeableness}: appreciative, forgiving, generous, kind, and sympathetic.
    \item \textbf{\underline{N}euroticism}: anxious, self-pitying, tense, touchy, unstable, and worrying.
\end{itemize}

In our research, we use the International Personality Item Pool (IPIP) and its IPIP-NEO derivations \cite{goldberg1999broad, goldberg2006international, johnson2005ascertaining, johnson2014measuring}, along with the BFI-S questionnaire by \cite{lang2011short} to provide the initial personality labels for participants. 
Each item in the questionnaire consists of a question and a set of options. All items are labeled with the corresponding Big Five personality factors annotated by psychologists for standardized personality assessment.

\subsection{Personality Traits Assessment}
Personality traits are crucial as they reflect long-term behavioral patterns with impacts on many life aspects \cite{caspi2005personality}. Studies show these traits significantly influence key life outcomes \cite{weiner2017handbook,caspi2005personality}, emphasizing the need for regular personality assessments in quality-of-life surveys.
Researchers have developed a variety of computational methods, employing diverse machine learning algorithms, to assess personality traits effectively.

\emph{Self-reported Personality Traits Assessment.}
Researchers have investigated whether self-reported personality traits can be evaluated accurately, \etc. For example, \cite{wolffhechel2014interpretation} utilized the Cubiks In-depth Personality Questionnaire (CIPQ 2.0), a normative self-reported tool, to measure various personality traits.

\emph{Peer Assessment.}
There is reliable psychological and biological evidence~\cite{kumari2004personality} suggesting that multimodal behaviors serve as reliable predictors of personality traits. These include audio signals~\cite{fang2016personality}, visual cues such as facial expressions and gestures~\cite{salam2016fully}, and observable social cues~\cite{principi2019effect}.

\emph{Computational Personality Traits Assessment.}
Humans can assess some personality traits from features, prompting research into whether computers can similarly automatically evaluate them.
For instance, face-image based approaches~\cite{ilmini2016persons,qin2018modern}, audio based approaches~\cite{gurpinar2016combining,zhang2016deep}, 
and face-to-face dyadic interaction-based scenarios~\cite{palmero2021context}.

\subsection{Personality Traits in Virtual Reality}

The importance of personality traits as factors significantly affecting the user experience in a virtual environment (VE) has been recognized in previous research, leading to several studies attempting to explore these effects~\cite{majumder2017deep,iacobelli2011large}. 
There has been research conducted on the effects of specific personality traits in VR environments, with studies dating back as far as 20 years. The personality traits most frequently studied 
in VR environments include absorption~\cite{tellegen1974openness}, mental imagination~\cite{sheehan1967shortened}, immersive tendencies~\cite{qin2009measuring}, locus of control~\cite{rotter1966generalized}, empathy~\cite{davis2018empathy}.
While this research highlights the need for further exploration of personality traits and their impact on the user experience in VR environments, such existing studies have primarily focused on presence~\cite{wallach2010personality,kober2013personality} 
and, to a lesser extent, on other aspects of the VR experience, such as the sense of embodiment\cite{dewez2019influence}.

Recent studies by \cite{katifori2022exploring} and \cite{maharani2023relationship} focus on task performance in VR and explore predicting user personality traits. Their research was limited to simple object manipulation tasks using hand movements. To our knowledge, There is no known research on the link between personality traits and performance in everyday activities in VR settings.

\section{\simulator Simulator}\label{sec:platform}

\subsection{Functions}
We develop an interactive and immersive VR simulator to stimulate cognitive progress, referred to as \simulator. This simulator is equipped with a range of advanced functionalities as follows.

\textbf{Daily Life Task Design.}
The primary goal of our simulator is to immerse users in everyday tasks that can reveal their personality traits. Since narrative elements can help to draw the users into the VR experience and make them more engaging, we integrate a role-playing narrative comprising a series of challenges. Participants engage in a "company onboarding" scenario, including riding an elevator with strangers, organizing a desk, and delivering a welcome speech, as depicted in Figure~\ref{fig:simulator}. These tasks were selected because of their relevance to human personality traits. In the elevator task, different personalities lead to varying choices in standing position, behavior in crowded elevators, and gaze direction~\cite{fryrear1976presses, rousi2020emotions}. Desk organization correlates with the "conscientiousness" trait in the Big Five personality model~\cite{freeman1990consciousness}, reflecting the degree of personal item orderliness. During public speaking, individuals with distinct personalities exhibit different performance indicators, including eye movements~\cite{kim2018aversive} and vocal responses, when they are observed by an audience, revealing their unique psychological states~\cite{kuai2021higher}.

Specifically, we instruct each participant to imagine themselves as a new employee about to join a multinational technology company and complete the onboarding process on the first day of employment with the company. The participant's initial task is to take an elevator ride with unknown colleagues and head to the office. Subsequently, they should gather office supplies and organize their desk. The final task involves going to the conference room to deliver a brief welcome speech to their colleagues. The instructions are presented to the participant at the beginning of each task.

\textbf{Interactive NPC.}
Believable NPCs of human behavior can enhance interactive applications across a variety of immersive environments~\cite{park2023generative}. To develop VR applications that are both immersive and engaging, we design NPCs capable of giving natural feedback to the user’s pose, touch, and voice commands. This design is based on multimodal large language models (MLLMs) that have been proven to understand human languages~\cite{vaswani2017attention,wei2023zero} and provide reactions similar to persons' personality~\cite{jiang2022mpi}, aiming to make interactions as natural and intuitive as possible, thereby enriching the overall VR experience.
The NPCs display natural body movements like fidgeting or gesturing during conversations, instead of staying static. They also have spontaneous conversations with each other. When touched by the user, NPCs turn to face them naturally. They act and respond like colleagues during discussions. If the participant is visible or talking to them, NPCs maintain eye contact by following the user's movements.

\textbf{Unanticipated Event Generation.}
According to existing research~\cite{farizi2019facial,craig2013understanding}, 
movements within VR environments tend to be involuntary, thus making conscious control of these movements a challenging endeavor. Micro-expressions are often regarded as indicators of truthfulness since individuals usually cannot suppress their spontaneous behaviors in situations that impose cognitive load or involve responding to unanticipated questions~\cite{depaulo2003cues}.

Fortunately, unanticipated events can be easily created by \simulator. Take the third task \textit{Organizing Your Desk} as an example, we design a mechanism that can occasionally and automatically disrupt the items on the office table without warning.
Based on our experimental observations, participants with high conscientiousness tend to make repeated efforts to organize items that have been disrupted. On the other hand, participants who do not have high conscientiousness but try to pretend to be conscientious are more likely to give up on tidying up.

\textbf{Objects.}
The environment in which our personality assessment occurs can greatly impact the level of immersion.  Consequently, we strive to create a visually appealing and interactive environment. For instance, objects within this environment can be grasped and moved. The elevator button can be pressed. When pressed, lights up to signal its activation, and the door opens.

\subsection{Usage}

\begin{figure}[]
    \centering
    \includegraphics[width=1\linewidth]{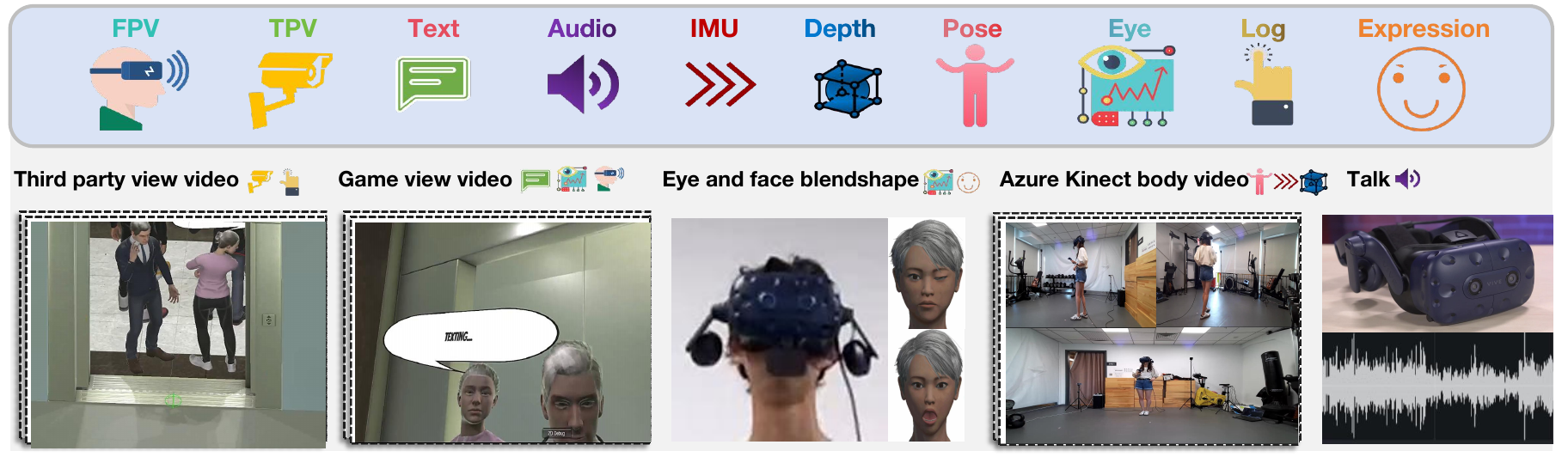}
   \caption{Illustration of collected dataset and modalities.}
    \label{fig:data}
        \vspace{-5mm}
\end{figure}

The recorded data includes multimodal information from participants who wear a head-mounted depth sensor, and a spatial microphone array enacting their roles in a simulated household environment, as shown in Figure~\ref{fig:data} and Figure~\ref{fig:simulator}.
Our experimental site is equipped with Azure Kinect DK, an integrated device featuring a depth sensor, spatial microphone array, video camera, and orientation sensor, complemented by comprehensive software development kits (SDKs).

The usage steps of \simulator are as follows:

    \textbf{\textit{Prepare the VR environment.}} The VR environment needs to be properly set up. Subsequently, the participant is required to wear the VR headset and be able to move around freely in the environment.
    
    \textbf{\textit{Introduce the experiment to participants.}} Each participant signs an informed consent form, 
    and is informed of an introduction to the experiment, details of the experimental procedures, information about potential risks, a privacy statement, compensation details, and a 10-minute exercise.
    
    \textbf{\textit{Start the assessment.}} Each participant is required to engage in three tasks, each lasting 10 minutes. Throughout the assessment, two managers will oversee the participants to ensure their comfort and safety. 
    Additionally, three psychology experts will observe the participants' behaviors and performance, and they will validate and revise the self-assessed answers to the OCEAN personality test completed by the participants at the time of registration.

\textbf{Equipment}
Participants are required to complete three tasks while wearing a VR headset, a facial expression tracker, and holding handsets as depicted in Figure~\ref{fig:simulator}.
We utilize the Vive Pro headset and Azure Kinect, which includes an RGB camera, depth sensor, spatial microphone array with a video camera, and orientation sensor as an all-in-one device with software development kits. As for the platform, we leverage the Unity game engine to develop immersive VR scenarios and execute these applications on the SteamVR platform.

\section{Dataset}\label{sec:data}

\subsection{Task Formulation}\label{sec:task_formulation}
We introduce a new benchmark for personality prediction by recording the multimodal data of selected participants as they complete tasks within \simulator.

\subsection{Participants Selection}
To recruit suitable participants, recruitment posters are disseminated through social media platforms and online forums, requiring each prospective participant to provide basic personal information, including gender and age, \etc. And complete an OCEAN personality traits questionnaire. When selecting participants, we place particular emphasis on balancing the gender ratio and personality traits among participants. We ensure a 1:1 male-to-female ratio and select participants based on their OCEAN personality questionnaire results, aiming to balance the representation of each trait within the participants. Furthermore, the study mainly involves individuals aged 18 to 35, who tend to adapt to VR technology quickly, which improves the quality of our dataset.

\subsection{Data Construction}\label{sec:data_construction}
We collect data from the selected participants, and three psychology experts review and revise the initial personality test through the observation of the participant's performance during interactions. The dataset after verification and correction is then utilized for training the personality prediction model.

\textbf{Data Collection.} 
The dataset comprises multimodal information recorded while participants perform tasks in \simulator, as shown in Figure~\ref{fig:data}. We collect data from 150 participants, including ten modalities such as first/third-person video, audio, text, eye tracking, facial microexpression, pose, depth data, log, and inertial measurement unit. Additionally, it includes information about positions, interaction with objects and NPCs, \etc. The human poses in each frame of the dataset are labeled following the COCO human pose style.

\textbf{Quality Control in Expert Review and Labeling.}
Considering the preliminary questionnaire results are based on participants' self-reports and reflect their subjective views of their personality traits, they are not completely accurate. To improve the dataset's reliability and accuracy, we have three psychology experts observe the participants during tasks and adjust their questionnaire results if needed. The final personality assessments are decided by majority vote among these experts, ensuring objectivity and accuracy in our dataset.

\textbf{Dataset Analysis.}\label{sec:data_analysis}
The detailed statistics of our collected dataset are shown in Table~\ref{tab:data_analysis}. The dataset comprises RGB+D video frames, the mesh of objects scene point clouds, action labels, and activity labels. This dataset encompasses data from 150 participants, consisting of 450 videos, 24.3 million RGB-D frames, and 20 motion categories. 
The total duration of recordings for all participants is over 4500 minutes, with a total number of 27000 × 5 clips (5 for the following video types including RGB video, depth image, audio, skeleton, and bounding box). 
We draw the bounding box of the object that the participant interacts with, along with hand localization.
The datasets have been segmented into 10-second clips, and divided into into training, development, and testing with a 0.6/0.2/0.2 ratio.

\begin{table}[h]
\begin{center} 
\caption{Statistics of multi-modal dataset.} 
\label{tab:data_analysis} 
 \resizebox{0.57\linewidth}{!}{
        \begin{tabular}{c|c}
            \hline \toprule
            \textbf{Modality}         & \textbf{Amount}  \\ \hline
            \text{RGB+D Body video}      & 450      \\
            \text{Multi-view video}       & 300    \\
            \text{Eye and face blendshape} & 150   \\
            \text{Audio}    & 150          \\
            \text{Text}     & 150          \\
            \text{Supplementary log}     & 600    \\
            \bottomrule
        \end{tabular}}%
\end{center} 
  \vspace{-7mm}
\end{table}

\section{Personality Traits Inference}\label{sec:experiments}
This section assesses the collected dataset's capacity to infer personality traits. We utilize a multi-modal transformer-based model~\cite{vaswani2017attention} to predict the personality traits of participants in \simulator.

In our research, we predict personality traits by translating human activities in a simulator into five OCEAN scores, treating each as a separate regression task. We use a transformer-based model to estimate these traits, converting observed behaviors within \simulator into quantifiable OCEAN personality scores.

The model comprises three modules: feature extraction, modality fusion, and OCEAN score prediction. 
The lower part of Figure~\ref{fig:motivation} provides a visual representation of all components and their interconnecting information flow. Detailed descriptions of these components are also presented below.

\begin{itemize}[leftmargin=*,noitemsep,nolistsep]
    \item \textbf{Feature extraction}: Human activities generate synchronous multi-modal data that can be recorded by \simulator. Following \cite{zhao2023mmicl}, different sources of context are captured in the form of different modalities including first/third-person view video, audio, text, eye tracking, facial microexpression, pose, depth data, Logs, and Inertial Measurement Unit (IMU). In the feature extraction module, we employ separate encoders for each modality. 
    We leverage the CLIP model~\cite{radford2021learning} for image and text encoder. We convert audio modality into 2D mel-spectrograms~\cite{gong2021ast}, and depth modalities into one-channel images and adopt ViT-S~\cite{tong2022videomae} encoders respectively. The body joint coordinates are obtained offline from Azure Kinect camera videos, and we address occlusion issues by calibrating the cameras and selecting unobstructed frames for data extraction. These joint coordinates are then encoded using a Multilayer Perceptron (MLP). For face and eye blend shapes, Temporal Convolutional Networks (TCN)~\cite{lea2017temporal} are utilized.

    \item \textbf{Modality fusion}: To obtain a fixed-size embedding that is both normalized and utilized in the InfoNCE loss~\cite{oord2018representation}, we incorporate a linear projection head specific to each modality into every encoder.
    All sensors are synchronized. We utilize the attention mechanism to model the relation between each pair of two modalities to obtain the ultimate input for the transformer layers.
    \item \textbf{OCEAN score prediction}
    We obtain OCEAN traits by applying a fully connected layer of the transformer. An Ocean score is computed for each participant.
\end{itemize}
The assessment of results is carried out based on the Mean Squared Error (MSE) between the aggregated personality trait score and the corresponding ground truth label (\ie, Expert-reviewed and verified OCEAN personality test results.) for each participant in the test set.

\section{Experiments and Results}\label{sec:results}
\textbf{Dataset Comparison}
We compare the datasets we collected with relevant personality research datasets. To the best of our knowledge, our dataset is the first to be collected within a VR environment, encompassing ten modalities for a comprehensive assessment of personality, setting it apart from existing datasets. See Table \ref{tab:datasets} for a numerical comparison.

The Cocktail Party dataset \cite{zen2010space} is a pioneering collection for recording six-person social interactions to analyze spatial and attention dynamics related to personality. CoffeeBreak \cite{cristani2011social} examines social group structures and orientations. IDIAP Poster \cite{hung2011detecting} offers overhead views for detecting F-Formations during presentations. Big Game \cite{hung2013classifying} focuses on identifying social actions in a quiz game context with sensor and video data. SALSA \cite{alameda2015salsa} integrates video, sensor information, and personality traits for social behavior analysis. MatchNMingle \cite{cabrera2018matchnmingle} documents interactions in speed-dating and parties with comprehensive behavioral annotations. UDIVA \cite{palmero2021context} provides a non-acted dyadic interaction dataset with audiovisual, physiological, and sociodemographic data, along with personality and relationship insights.

\begin{table}[h]
\begin{center} 
\caption{Comparison of Personality research datasets.}
     \label{tab:datasets}
    \small
    \scalebox{0.9}{
  \begin{tabular}{lcccccccccc}
\toprule
\multirow{2}{*}{\thead{Dataset}} & \multirow{2}{*}{\thead{Numb.\\ of\\ people}} & \multirow{2}{*}{\thead{Total\\ time\\ (minutes)}} & \multirow{2}{*}{\thead{VR}} & \multicolumn{5}{c}{\thead{Modalities}} \\
    \cmidrule(l){5-9}
    & & & & \textbf{V} &  \textbf{A} & \thead{F} & \textbf{P}  & \textbf{I} \\
    \midrule
    Cocktail & 6 & 30 & \ding{55} & \ding{51} & \ding{55} & \ding{55} & \ding{55} & \ding{55}\\
    CoffeeBreak  & 10 & - & \ding{55} & \ding{51} & \ding{55} & \ding{55} & \ding{55} & \ding{55} \\
    Big Game  & 32 & 30 & \ding{55} & \ding{51} & \ding{55} & \ding{55} & \ding{55} & \ding{55}\\
    Idiap  & 50 & 360 & \ding{55} & \ding{51} & \ding{55} &  \ding{55} & \ding{55} & \ding{55}\\
    SALSA  & 18 & 60 & \ding{55} & \ding{51} & \ding{51} & \ding{55} & \ding{55} & \ding{51}\\
    MatchNMingle & 92 & 120 & \ding{55} & \ding{51} & \ding{51} & \ding{55} & \ding{55} & \ding{51}\\
    UDIVA & \underline{147} & \textbf{5430} & \ding{55} & \ding{51} & \ding{51} &\ding{55} & \ding{55} & \ding{51} \\ 
    \midrule
    \textbf{Ours} & \textbf{150} & \underline{4500} & \ding{51} & \ding{51} & \ding{51} & \ding{51} & \ding{51} & \ding{51} \\ 
    \bottomrule[1pt]
  \end{tabular}
      }
  \end{center}
  \vspace{-9mm}
\end{table}

\begin{table}[h]
\begin{center} 
\caption{Accuracy results of estimating the personality traits. The higher, the better.}
\label{tab:all}
\scalebox{0.75}{
\begin{tabular}{lcccccc}
\toprule[1pt]
\multicolumn{1}{c}{\textbf{Methods}} & \multicolumn{1}{c}{\begin{tabular}[c]{@{}c@{}}\textbf{O}\end{tabular}} & \multicolumn{1}{c}{\begin{tabular}[c]{@{}c@{}}\textbf{C}\end{tabular}} &\multicolumn{1}{c}{\begin{tabular}[c]{@{}c@{}}\textbf{E}\end{tabular}} & \multicolumn{1}{c}{\begin{tabular}[c]{@{}c@{}}\textbf{A}\end{tabular}} & \multicolumn{1}{c}{\begin{tabular}[c]{@{}c@{}}\textbf{N}\end{tabular}} & \multicolumn{1}{c}{\begin{tabular}[c]{@{}c@{}}\textbf{All}\end{tabular}} \\ \midrule[1pt]
FGM-UTRECHT & 70.36  & 65.22   & 73.98     & 60.46  & 56.14     & 65.23           \\
SMART-SAIR & \underline{72.65}   & \textbf{67.02}   & \underline{74.52}     & \underline{64.93}   & \underline{57.32}    & \underline{67.29}           \\
\midrule
\textbf{Ours} & \textbf{75.43}  & \underline{66.96}  & \textbf{75.93}  & \textbf{68.03}  & \textbf{60.87}    & \textbf{69.44}                                                         \\ 
\bottomrule[1pt]
\end{tabular}
}
\end{center} 
  \vspace{-2mm}
\end{table}

\textbf{Baselines} Due to the scarcity of work utilizing multimodal data for personality prediction, we compare our method with SMART-SAIR~\cite{salam2021fact} and FGM-UTRECHT~\cite{pessanha2022fact}, two works in the ICCV 2021 Understanding Social Behavior Challenge \cite{palmero2021context}.
Building on studies about gender-specific 
personality traits~\cite{weisberg2011gender}, SMART-SAIR separates participants by gender and crafted a gender-specific multimodal model using Neural Architecture Search~\cite{jin2019auto}, integrating visual and textual data for personality prediction.
FGM UTRECHT develops a Random Forest (RF) regressor using only metadata features like age, gender, and session count. They also combined linguistic, audio, and metadata features in a multitask approach using RF regressors and late fusion.

\paragraph{Estimate Personality Traits Using \simulator.}
We compare our method with SMART-SAIR and FGM-UTRECHT.
As shown in Tables~\ref{tab:mse} and Table~\ref{tab:all}. From Table~\ref{tab:mse}, Notably, our approach achieved the lowest mean square error for the Openness score, followed by the Extraversion score, outperforming the two baselines. This suggests that our method provides better estimates of self-assessed personality traits. Table~\ref{tab:all} presents a summary of the accuracy comparison.

\begin{table}[!ht]
\begin{center} 
\caption{MSE results of estimating the personality traits. The lower, the better.} 
\label{tab:mse}
\scalebox{0.75}{
\begin{tabular}{lcccccc}
\toprule[1pt]
\multicolumn{1}{c}{\textbf{Methods}} & \multicolumn{1}{c}{\begin{tabular}[c]{@{}c@{}}\textbf{O}\end{tabular}} & \multicolumn{1}{c}{\begin{tabular}[c]{@{}c@{}}\textbf{C}\end{tabular}} &\multicolumn{1}{c}{\begin{tabular}[c]{@{}c@{}}\textbf{E}\end{tabular}} & \multicolumn{1}{c}{\begin{tabular}[c]{@{}c@{}}\textbf{A}\end{tabular}} & \multicolumn{1}{c}{\begin{tabular}[c]{@{}c@{}}\textbf{N}\end{tabular}} & \multicolumn{1}{c}{\begin{tabular}[c]{@{}c@{}}\textbf{All}\end{tabular}} \\ \midrule[1pt]
FGM-UTRECHT & 0.9640   & 0.9147   & 0.9852     & 0.9039   & 1.0033      & 0.9542           \\
SMART-SAIR & \underline{0.8434}  & \textbf{0.8302}   & \underline{0.8894}     & \underline{0.8823}    & \underline{0.9071}     & \underline{0.8705}           \\
\midrule 
\textbf{Ours}   & \textbf{0.8112}  & \underline{0.8554}  & \textbf{0.8298}  & \textbf{0.8801}  & \textbf{0.8839}    & \textbf{0.8521}                                                         \\ \bottomrule[1pt]
\end{tabular}
}
\end{center} 
  \vspace{-2mm}
\end{table}

\textbf{Impact of Modalities.}
We conduct an ablation study to demonstrate the influence of various modalities. 
As shown in Table~\ref{tab:Ablation study}, the removal of modalities results in a further reduction of prediction capability. These findings underscore the significant role that different modalities play.

\begin{table}[h]
\begin{center} 
\caption{Accuracy of estimating personality traits. We analyze the impact of different modalities. The higher, the better. A-Audio, T-Text, P-Pose, F-Face Blendshape, I-Interaction.}

\label{tab:Ablation study}
\scalebox{0.9}{
\begin{tabular}{lcccccc}
\toprule[1pt]
\multicolumn{1}{c}{\textbf{Modality}} & \multicolumn{1}{c}{\begin{tabular}[c]{@{}c@{}}\textbf{O}\end{tabular}} & \multicolumn{1}{c}{\begin{tabular}[c]{@{}c@{}}\textbf{C}\end{tabular}} &\multicolumn{1}{c}{\begin{tabular}[c]{@{}c@{}}\textbf{E}\end{tabular}} & \multicolumn{1}{c}{\begin{tabular}[c]{@{}c@{}}\textbf{A}\end{tabular}} & \multicolumn{1}{c}{\begin{tabular}[c]{@{}c@{}}\textbf{N}\end{tabular}} & \multicolumn{1}{c}{\begin{tabular}[c]{@{}c@{}}\textbf{All}\end{tabular}} \\ \midrule[1pt]
A        & 58.23  & 54.02    & 56.96    & 57.39   & 52.27   & 55.77             \\
T        & 56.83  & 52.96    & 50.49    & 57.93   & 56.02   & 54.85                 \\
P        & 54.20  & 53.87    & 55.77    & 53.20   & 52.86   & 53.98              \\
F        & 55.79  & 56.04    & 58.31    & 53.98   & 54.28   & 55.68       \\
A-T      & 65.39  & 61.10    & 66.99    & 62.83   & 58.24   & 62.91      \\
A-T-P    & 67.29  & 63.89    & 70.65    & 64.20   & 60.98   & 65.40       \\
A-T-P-F  & 70.33  & 64.71    & 72.73    & 65.43   & \textbf{61.26}    & 66.89  \\
\midrule
\textbf{A-T-P-F-I}  & \textbf{75.43}  & \textbf{66.96}  & \textbf{75.93}  & \textbf{68.03}  & 60.87    & \textbf{69.44} \\
\bottomrule[1pt]
\end{tabular}
}
\end{center} 
    \vspace{-7mm}

\end{table}

\section{Conclusion}
We develop a VR simulator to immerse the participant in the VR experience and make it more engaging to reveal their personality. With this simulator, we collect a large-scale, synchronous, multi-modal dataset with ten modalities. Furthermore, we conduct extensive research on different types of modalities and demonstrate the superiority of \simulator.

\section{Acknowledgments}

We express our gratitude to the anonymous reviewers for their valuable comments. Special thanks to BIGAI for providing equipment and support for the experimental site, and heartfelt appreciation to every participant who signed up. The work is supported by the National Science and Technology Major Project of China (2022ZD0114900), the Talent Fund of Beijing Jiaotong University (2023XKRC006) to WH, the National Natural Science Foundation of China
(32200854) and the Young Elite Scientists Sponsorship Program
(2021QNRC00) to YP.

\nocite{ChalnickBillman1988a}
\nocite{Feigenbaum1963a}
\nocite{Hill1983a}
\nocite{OhlssonLangley1985a}
\nocite{Matlock2001}
\nocite{NewellSimon1972a}
\nocite{ShragerLangley1990a}

\bibliographystyle{apacite}

\setlength{\bibleftmargin}{.125in}
\setlength{\bibindent}{-\bibleftmargin}

\bibliography{CogSci_Template}

\end{document}